\documentclass[twocolumn,prb,showpacs,preprintnumbers,superscriptaddress,amsmath,amssymb]{revtex4}

\usepackage{graphicx}
\usepackage{dcolumn}
\usepackage{bm}
\begin{document}
%
%\preprint{APS/123-QED}
%
\title{Remarkable change of tunneling conductance in YBCO films in fields up to 32.4T}
\author{R. Beck}
\email{roy@post.tau.ac.il}

\affiliation{School of Physics and Astronomy, Raymond and Beverly
Sackler Faculty of Exact Science, Tel-Aviv University, 69978
Tel-Aviv, Israel.}
\author{Y. Dagan}
\affiliation{School of Physics and Astronomy, Raymond and Beverly
Sackler Faculty of Exact Science, Tel-Aviv University, 69978
Tel-Aviv, Israel.} \affiliation{Center for Superconductivity
Research, Department of Physics, University of Maryland, College
Park, Maryland 20742, USA.}
\author{A. Milner}
\altaffiliation{Current address: Department of Chemical Physics,
Weizmann Institute of Science, Rehovot 76100, Israel.}
\affiliation{School of Physics and Astronomy, Raymond and Beverly
Sackler Faculty of Exact Science, Tel-Aviv University, 69978
Tel-Aviv, Israel.}

\author{G. Leibovitch}
\affiliation{School of Physics and Astronomy, Raymond and Beverly
Sackler Faculty of Exact Science, Tel-Aviv University, 69978
Tel-Aviv, Israel.}
\author{A. Gerber}
\affiliation{School of Physics and Astronomy, Raymond and Beverly
Sackler Faculty of Exact Science, Tel-Aviv University, 69978
Tel-Aviv, Israel.}
\author{R. G. Mints}
\affiliation{School of Physics and Astronomy, Raymond and Beverly
Sackler Faculty of Exact Science, Tel-Aviv University, 69978
Tel-Aviv, Israel.}
\author{G. Deutscher}
\affiliation{School of Physics and Astronomy, Raymond and Beverly
Sackler Faculty of Exact Science, Tel-Aviv University, 69978
Tel-Aviv, Israel.}
\date{\today}
\begin{abstract}
We studied the tunneling density of states in YBCO films under
strong currents flowing along node directions. The currents were
induced by fields of up to 32.4T parallel to the film surface and
perpendicular to the $CuO_{2}$ planes. We observed a remarkable
change in the tunneling conductance at high fields where the
gap-like feature shifts discontinuously from 15meV to a lower bias
of 11meV, becoming more pronounced as the field increases. The
effect takes place in increasing fields around 9T and the
transition back to the initial state occurs around 5T in
decreasing fields.
\end{abstract}
\maketitle
\section{Introduction}

The order parameter of a \emph{d}-wave superconductor has
node-lines located at angles $\theta_\pm =\pm\pi/4$, where
$\theta$ is the angle between the quasiparticle momentum and the
crystallographic [1,0,0] direction \cite{d-wave}. As a result, the
tunneling density of states of a \emph{d}-wave superconductor is
significantly different from that of a conventional \emph{s}-wave
superconductor. In particular, it reveals the existence of low
energy surface bound states, which are the origin of the zero bias
conductance peak at pair breaking surfaces \cite{hu1,tanaka2}. The
high conductance at low bias, below the \emph{d}-wave gap, is in
sharp contrast with the low conductance in an \emph{s}-wave
superconductor at similar bias. The \emph{d}-wave gap itself is
marked in the tunneling density of states by a weak structure
called the gap-like feature \cite{tanaka2} (see Fig. 1). The zero
bias conductance peak and gap-like feature are well identified in
the tunneling density of states of high-$T_c$ cuprates
\cite{covington3,dagan6} and simultaneously observed when the
surface roughness scale is smaller than the junction size
\cite{dagan6,FSR15}.
\begin{figure}
\includegraphics[width=0.72\hsize]{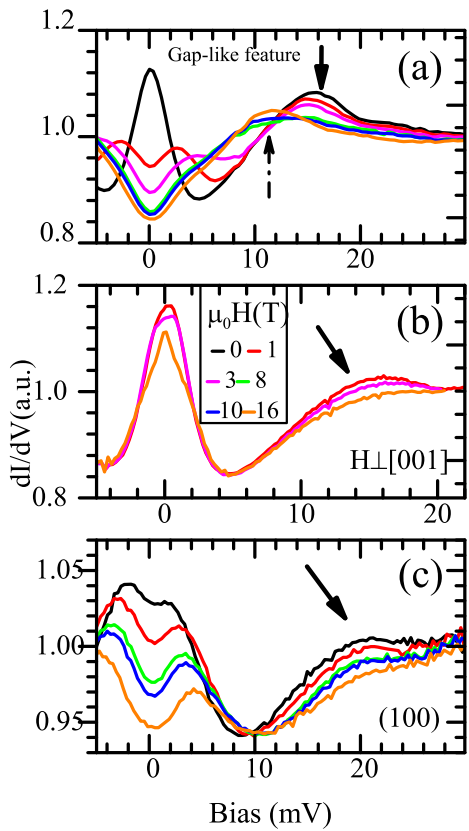}
\caption{\label{fig:f1} dI/dV versus bias voltage; magnetic field
applied parallel to the films surface up to 16T at 4.2K. (a) A
(110) in-plane orientated film at increasing field. The field is
perpendicular to the $CuO_2$ planes (sample 1).(b) A (110)
in-plane orientated film at increasing field. The field is
parallel to the $CuO_2$ planes (sample 1). (c) A (100) in-plane
orientated film.  The field is perpendicular to the $CuO_2$
planes. Solid (dashed) arrows indicate the gap like feature
positions at low (high) fields. Note that only when the field is
perpendicular to the $CuO_2$ planes and the normal to the film
surface is a nodal direction, a remarkable change in the spectrum
is observed at high fields. }
\end{figure}
It was predicted that a \emph{d}-wave symmetry can be modified by
a perturbation that creates a gradient of the order parameter.
This is the case  of a vortex core \cite{tesanovic}, sample
surface \cite{tanuma2001}, and currents
\cite{khavkine04,zapotosky}.
\par
In this study we report measurements of the conductivity of
In/YBCO junctions. Currents in the YBCO film are induced by
applying magnetic fields, parallel to the surface and
perpendicular to the $CuO_2$ planes. Films having (110) and (100)
orientation are used respectively to induce nodal and anti-nodal
currents.
\par
For the (110) orientation the tunneling conductance changes
remarkably in high magnetic fields - high currents in a domain
that has not been investigated until now, with applied fields
reaching up to 32.4T. The position of the gap-like feature shifts
down discontinuously in increasing fields around 9T and in
decreasing fields around 5T. We argue that these shifts are due to
nodal surface currents induced by the applied field, with the
field itself, possibly inducing a certain modification of the
vortex state. No transition is observed when the field is parallel
to the $CuO_2$ planes (Fig. 1b) or when the film has the (100)
orientation (Fig. 1c). In both cases there are no currents flowing
along the nodal direction.
\par
\section{Experimental results}
Our oriented films were sputtered onto (110) $SrTiO_{3}$ and (100)
$LaSrGaO_{4}$. All films have a critical temperature of 89K
(slightly underdoped). Tunneling junctions were prepared by
pressing a freshly cut Indium pad onto the films' surface
\cite{dagan8}. These junctions are of high quality, as shown by
their low zero bias conductances below the critical temperature of
the Indium electrodes \cite{dagan8}. A schematic drawing of the
crystallographic orientation and experimental setup is shown in
Fig. 2.
\par
\begin{figure}
\includegraphics[width=0.70\hsize]{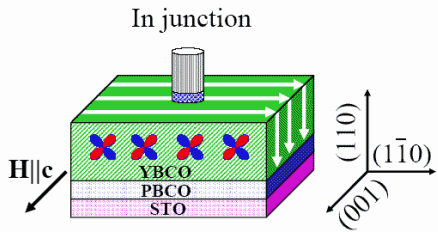}
\caption{\label{fig:f0} Schematic presentation of the measurement
setup for the (110) films. Indium pads are pressed against the
surface of the oriented thin film. The orientation of the film
enables us to apply a magnetic field parallel or perpendicular to
the $CuO_2$ layers while the field is kept parallel to the films'
surface and perpendicular to the tunneling current.}
\end{figure}
\par
Tunneling characteristics in (110) films at zero magnetic field
exhibit the known zero bias conductance peak and gap-like feature
(Fig. 1a). The magnetic field splitting of the zero bias
conductance peak was previously addressed
\cite{covington3,dagan6,dagan8,FSR15,beck16,beck20}. We focus here
on the field dependence of the gap-like feature at high bias. It
shows a progressive, roughly linear, shift of the peak position
from ~17 meV to ~15 meV as the field increases from 0 to 1T. This
initial decrease is followed by a flat region up to 6T. If that
field is not exceed and then reversed a hysteresis loop is
described ending up with a flat region at low field. If the field
is increased above 6T the gap-like feature amplitude starts to
shrink until, at 8T, it can't be identified anymore. In the range
of 8 to 11T a flat maximum develops between 10 and 15 meV. An 11
meV peak builds up with the field and is clearly identified above
11T. Up to a field of 16T, no detectable smearing of this peak
occurs.
\par
Reducing the field from values larger than 11T has another
interesting effect. The 11 meV peak gradually shifts back to 14
meV as the field is reduced by about 1T, for example from 15 to
14T (Fig. 3) or from 32.4 to 31.5T (Fig. 4b). In contrast to the
9T field up transition, the shift back to 14 meV is
\emph{continuous}, which shows that the new peak at 11 meV is
indeed a new gap like feature rather than being related for
instance to the split zero-bias conductance peak. By further
reducing the field, the 14 meV peak shrinks while the 16 meV
builds up below 5T (Fig. 5a). An analogous behavior is seen under
field cooled conditions (Fig. 5b).
\par
The overall variation of the gap-like feature peak position with
respect to the applied field can be seen in Fig. 3. The jump in
its position can be clearly observed in increasing fields above 8T
and decreasing fields lower than 6T. The gradual increase of the
11meV peak amplitude as the field is increased beyond 10T (Fig. 6)
suggests that it characterizes a new superconducting state.

\begin{figure}
\includegraphics[width=0.72\hsize]{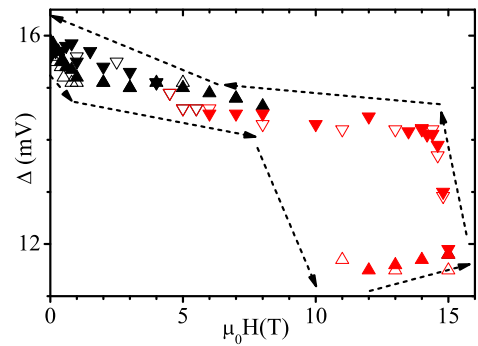}
\caption{\label{fig:f3} Gap-like feature position for sample 1 in
increasing $(\triangle)$ and decreasing $(\nabla)$ magnetic fields
at 4.2K. Data taken both for positive (full) and negative (hollow)
field polarity.Black (red) represent the low (high) field state}
\end{figure}

\section{Discussion}
The rapid change in the peak position upon field reversal (Figs.
3, 4b) means that its position is affected by field induced
currents. The strong difference between (100) and (110) oriented
films, both showing a similar gap-like feature, presumably due to
surface roughness \cite{FSR15}, indicates that these currents flow
over a depth much larger than the surface roughness (few tens of
nm).
\par
The Bean-Livingston barrier \cite{clem10} can prevent the entry of
vortices up to fields of the order of the thermodynamical critical
field, $H_{c}$ ($\sim 1$T for YBCO). The rapid initial decrease of
the gap-like feature peak position from 16 meV down to 14 meV over
that field range can be due to the delayed vortex entrance (see
Fig. 3). We name the corresponding currents surface vortex
currents, $j_{V}$. This is confirmed by the low field hysteresis
loop, as there is no Bean-Livingston barrier in decreasing fields.
This initial decrease is not observed in (100) oriented films
where the currents at the surface flow along an anti-nodal
direction. This decrease is therefore clearly due to nodal
currents. The major question raised here concerns however the
origin of the 11 meV peak seen in increasing fields above 10T. It
could be a current effect, or a field effect, or a combination of
both.
\par
In addition to the vortex surface currents, $j_V$, one should also
consider The Meissner screening current, $j_{M}$, and the Bean
current $j_{B}$ due to vortices pinned in the bulk. We showed
\cite{beck20} that by measuring the difference between field
cooled and decreasing field splitting values of the zero bias
conductance peak (also due to the surface currents \cite{FSR15}),
one can estimate the Bean critical current value. We found that
$j_{B}$ is roughly constant up to fields of 16T and has a value of
a few tens of MA/cm$^2$.
\par
\begin{figure}
\includegraphics[width=0.65\hsize]{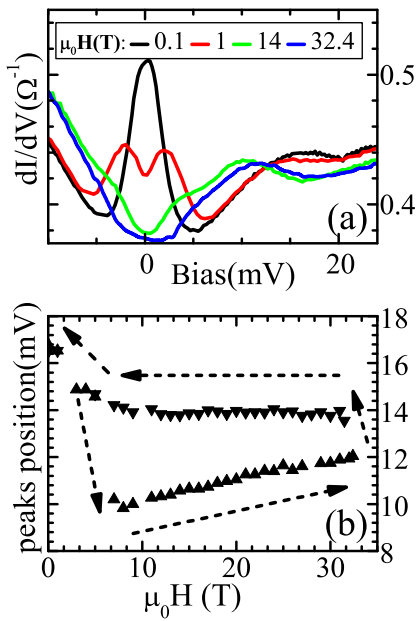}
\caption{\label{fig:f4} (a) dI/dV versus bias voltage for sample 2
measured at 1.3K in increasing magnetic fields. (b) gap-like
feature position in increasing $(\triangle)$ and decreasing
$(\nabla)$ fields.}
\end{figure}
The surface current can be obtained by calculating the depth, $d$,
of the vortex-core free area at the sample surface \cite{clem10}.
Its derivation is not affected by the \emph{d}-wave symmetry and
has to include the Bean current $j_B$. In the following the effect
of the vortex surface current is neglected. Consider a
semi-infinite superconductor in a uniform magnetic field $H$. The
field inside, $b(x)$, is the solution of London's equation which
has to match the boundary conditions $b(0)=H$, $b(d)=\tilde{B}$
and the vortex matter equilibrium condition $j(d)=j_B$, where
$\tilde{B}$ is the local induction value. In the field range
$H_{c1}\ll H \ll H_{c2}$ we have $d\ll\lambda$ and $j_B\ll j_M$:
\begin{equation}\label{d}
d\approx\lambda\sqrt{{2(H-\tilde{B})/H}}+4\pi j_B\lambda^2/cH,
\end{equation}
where $\lambda$ is London's penetration depth. The same
approximation results in $\tilde{B}\approx B$, where $B$ is the
equilibrium induction, $j\approx j_M(H)+j_B$ in increasing fields
and $j\approx j_M(H)-j_B$ in decreasing fields, where:
\begin{equation}
j_M=\frac{c}{4\pi\lambda}\sqrt{-8\pi HM}=
\frac{c}{4\pi\lambda^{2}}\sqrt{\frac{\phi_{0}H}{4\pi}\ln\frac{H_{c2}}{H}},
\end{equation}
and $\phi_{0}$ is the flux quantum. We find $j_M\sim 2.2\cdot
10^8\,$A/cm$^2$ for $H=90\,$KOe, $\lambda = 1500\AA$ and
$H_{c2}=1,200$KOe.
\par
We emphasize that the contributions of $j_V$, $j_M$ and $j_B$ may
all be important for the interpretation of the experiment. The
field reversal effect can be explained by $j_B$ and/or $j_V$.
After field reversal, $j_B$ changes sign while $j_V$ is negligible
\cite{clem10}.
\par
\begin{figure}
\includegraphics[width=0.85\hsize]{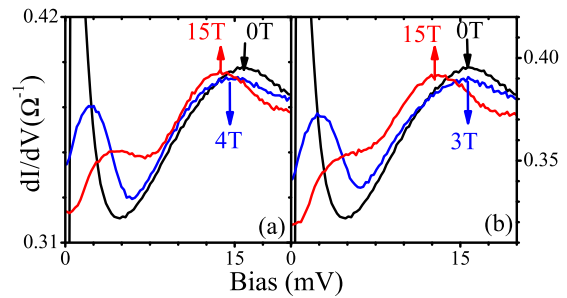}
\caption{\label{fig:5}  dI/dV versus bias voltage for sample 3
measured at 0.5K. (a) Decreasing fields from 16T. (b) Field cooled
conditions. Note that the intermediate field peak conductance is
lower than both the high and low field ones.}
\end{figure}
\par
We note that the progressive enhancement of the 11meV peak in
increasing fields (Fig. 6) suggests a transition to a new
superconducting state. Any continuous reduction of the
\emph{d}-wave gap would not be accompanied by an enhancement of
its gap-like feature peak amplitude.
\par
The new superconducting phase could be a different vortex state
\cite{maki2}, in such case the transition would be basically field
induced. Alternatively, the new phase could appear due to strong
nodal currents and possibly have an order parameter with a
symmetry different from a pure $d-wave$
\cite{khavkine04,zapotosky}.
\par
A general difficulty in comparing our data to existing theories is
that they have addressed only the small current limit
\cite{khavkine04,zapotosky,xu18}. We can only offer some
speculations as to what a high current phase might be. In a
previous publication \cite{beck16} we discussed the zero bias
conductance peak field splitting in terms of a field-induced
$id_{xy}$ component. But we have found no correlation between the
zero bias conduction peak splitting and the gap like feature
position implying that their origins are different. For instance,
after decreasing the field from 16 to 15T, the position of the
gap-like feature remains unchanged down to 5T (see Fig. 3), but
the zero-bias conductance peak splitting reduce from 4.2 to 2.5
meV (see Fig.4 in Ref. \cite{beck16}). We speculate that surface
currents on the coherence length scale could split the zero bias
peak, \cite{FSR15} but, as shown here, only currents on much
larger length scale are affecting the position of the gap like
feature position.
\par
The fact that we observe a regime where two gap features coexist,
both in increasing and decreasing fields as well as in field
cooled conditions, with a definite hysteresis, suggests a first
order transition showing superheating and supercooling effects. To
be specific, following the position and amplitude of the gap-like
feature as a function of applied field we observed that the
transition from low to high fields state takes place at 9T
(superheating) and back from the high to low fields at 5T
(supercooling).
\par
Whatever the high current - high field phase is, it is clear that
it has the effect of reducing the density of low lying states, and
this well beyond the field-current where the zero bias conductance
peak has disappeared (Fig. 4a). A transition to an inhomogeneous
state in the case of nodal currents was recently speculated about
by Khavkine \emph{et. al.} \cite{khavkine04}. We also note that in
very high fields the vortices nearest to the surface are located
within a few coherence lengths, which may affect the tunneling
conductance \cite{schophl21}.
\par
A different explanation to the data would be that the remarkable
change in the tunneling conductance at 9T in increasing fields is
a vortex state transition e.g. Bragg to votex glass
\cite{yeshurun}. Such a transition is known to be irreversible as
observed. The new vortex state could modify the pinning and hence
the total nodal current or its' effect on the order parameter.
This could explain the large hysteresis observed at high fields.
However, in contradiction to the high bias region, the low bias
region which is also sensitive to total current via the Doppler
shift mechanism, \cite{FSR15} does not show substantial
hysteresis.
\par

\begin{figure}
\includegraphics[width=0.65\hsize]{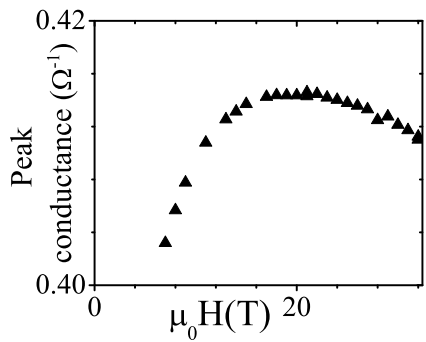}
\caption{\label{fig:6} 11 meV gap-like feature peak amplitude of
sample 2 for increasing fields. The enhancement up to 20T suggests
that the high fields state has a stronger coherence peak.}
\end{figure}
\section{Summary}
In conclusion, we have observed a remarkable change in the
tunneling conductance in high magnetic fields in YBCO (110)
oriented films. A transition of the gap-like feature position and
amplitude is present in both increasing and decreasing fields and
under field cooled conditions for fields oriented parallel to the
surface and perpendicular to the $CuO_2$ planes in such a way as
to induce currents along nodal directions. We have proposed that
the observed transition may be induced by these currents. In the
high current state, the density of low energy states is reduced,
possibly indicating the emergence of a component of the order
parameter leading towards a fully gaped state. Alternatively, the
changes in the tunneling characteristics may be due to a
transition between two vortex states, having different gap values
and sensitivity to nodal currents.
\par
This work was supported by the Heinrich Herz-Minerva Center for
High Temperature Superconductivity, the Israel Science Foundation,
the Oren Family Chair of Experimental Solid State Physics and NSF
grant DMR 01-02350. Work carried out at the National High Magnetic
Field Laboratory at Tallahassee is supported by an NSF cooperative
agreement DMR-00-84173 and the State of Florida. The work of R.M.
was supported in part by grant No.~2000011 from the United States
-- Israel Binational Science Foundation (BSF), Jerusalem, Israel.
G.D. is indebted to Philippe Nozieres for very helpful
discussions. We are indebted to Amlan Biswas (UF) for his
contribution to the NHMFL experiment and to Amir Kohen for
discussions.
\par

\end{document}